\documentclass[]{article}
\usepackage{amsmath}
\usepackage{amsfonts}
\usepackage{amssymb}
\usepackage{graphics,graphicx}
\usepackage{cite}
\usepackage{latexsym}
\usepackage{ulem}
 
\begin{document}
\title{Interpenetrating subspaces \\ as a funnel to extra space}
\author{Sergey G. Rubin \\
National Research Nuclear University "MEPhI"\,, \\ (Moscow Engineering Physics Institute)  \\
sergeirubin@list.ru}

\maketitle

\begin{abstract}
New solution for two interpenetrating universes is found. Higher derivative gravity acting in 6-dimensional space is the basis of the study that allows to obtain stable solution without introducing matter of any sort. Stability of the solution is maintained by a difference between asymptotic behavior at spacial infinities. For an external observer such a funnel looks similar to a spherical wormhole.
\end{abstract}

\section{Introduction}

The theory of gravity provides us with great variety of nontrivial objects. Most known of them are black holes and wormholes. The existence of black holes acquires substantial observational background nowadays. Wormholes are considered as a hypothetical way to pass from one large space to another: a property which distinguishes traversable wormholes from black holes \cite{Bron}, \cite{ArmendarizPicon}. Other stable topological configurations of space (geons) are considered as particle-like objects possessing a mass and a charge. The problem of a topology changes is discussed in \cite{deAlbuquerque}.
A Lorentzian wormhole is a space-time whose spatial sections contain
two asymptotically flat regions joined by a "throat". Most of the literature is devoted to 4-dim wormholes, however 
there is substantial branch concerning spaces of an arbitrary dimensionality, see e.g. \cite{DeBenedictis}. 
Lorentzian wormholes embedded in the de Sitter space are discussed in \cite{Roman}, \cite{Krishnan}, \cite{Sushkov}.
 
In this paper, new solution with properties similar to those of a wormhole is discussed. The result is based on the extra space idea with a dimensionality $D>5$ and the gravity with higher derivatives.  Multidimensional wormholes are also  discussed in the literature see e.g. \cite{Dzhun}.  The interest in $f(R)$ theories is motivated by inflationary scenarios starting from the pioneering work of Starobinsky \cite{Star80}. A number of viable $f(R)$ models in 4-dim space that satisfy the observable constraints are proposed in Refs. \cite{Amendola}, \cite{Starob1}, \cite{Odin}. Also, substantially new results may be obtained on the basis of $f(R)$-theories of gravity, see \cite{BOOK}  \cite{DiCriscienzo} and references therein.

In this paper it is supposed that our Universe is described by a $D$-dim space ($D>4$) with a topology $T\times \mathbb{V}_{D-1}$. Its volume was comparable with unity in the Planck units at the moment of its origination from the space time foam. In the following three of these dimensions grew while others remained compact and/or small.
It seems reasonable to suppose that the choice was made accidentally depending on initial conditions, see e.g. \cite{ab(t)1}. 
It is commonly accepted that our Universe belongs to only one of such 3-dim subspaces. If it is not true new space geometry caused by nontrivial boundary conditions at infinity is formed. Its structure is studied below. 
For an external observer the solution looks like a wormhole though its internal geometry is different.

\section{Boundary conditions \\ and funnel geometry}

Let us start with metric
\begin{equation}\label{metric0}
ds^2 =  G_{\mu\nu}dZ^{\mu}dZ^{\nu}+G_{ab}dZ^adZ^b
\end{equation}
A lot of literature (see \cite{KKReview} for review) is devoted to study this metric.
One of the simplest geometry described by metric \eqref{metric0} is the direct products $M_4 \times V_{D-4}$ of the Minkowski space and a $(D-4)$-dim extra space with metric
\begin{eqnarray}\label{sol0}
&&G_{\mu\nu}=diag(1,-1,-Z_2^2,-Z_2^2\sin^2Z_3),  \\
&&G_{ab}=r_0 \cdot diag(-1,-\sin^2 Z_6,-\sin^2 Z_6\sin^2 Z_7,...),\nonumber\\
&&\mu,\nu=1,2,3,4\quad a,b=5,6,...,D,\nonumber
\end{eqnarray}
where $r_0$ is a radius of $(D-4)$-dim sphere with coordinates  $Z_5, ... ,Z_D$. The coordinates $Z_1, Z_2, Z_3, Z_4, (-\infty < Z_2 <\infty , 0<Z_3<\pi, 0<Z_4<2\pi )$ describe the extended Minkowski space with the Ricci scalar $R_4 = 0$. 

Let us consider more interesting case with the 4-dim metric depending on a single coordinate $Z_2$. It holds if boundary conditions at $Z_2 \rightarrow +\infty$ and  at $Z_2 \rightarrow -\infty$ differ from each other. More definitely, suppose that first condition at $Z_2 \rightarrow +\infty$ coincides with static geometry \eqref{sol0}.
The subspace described by space coordinates $2,3,4$ is assumed to be large.

Another boundary condition  at $Z_2 \rightarrow -\infty$
\begin{eqnarray}\label{Bound2}
&&G_{\mu\nu}(Z_2\rightarrow -\infty)=diag(1,-1,-Z_2^2,-Z_2^2\sin^2Z_6),\quad  \\
&&G_{ab}(Z_2\rightarrow -\infty)=r_0 \cdot diag(-1,-\sin^2 Z_3,-\sin^2 Z_3\sin^2 Z_4,...),\nonumber\\
&&\mu,\nu=1,2,6,7\quad a,b=5,3,4,8...,D.\nonumber
\end{eqnarray}
relates to another static large subspace with space coordinates $2,6,7$. 
\begin{figure}
	\centering
	\includegraphics[width=0.8\linewidth]{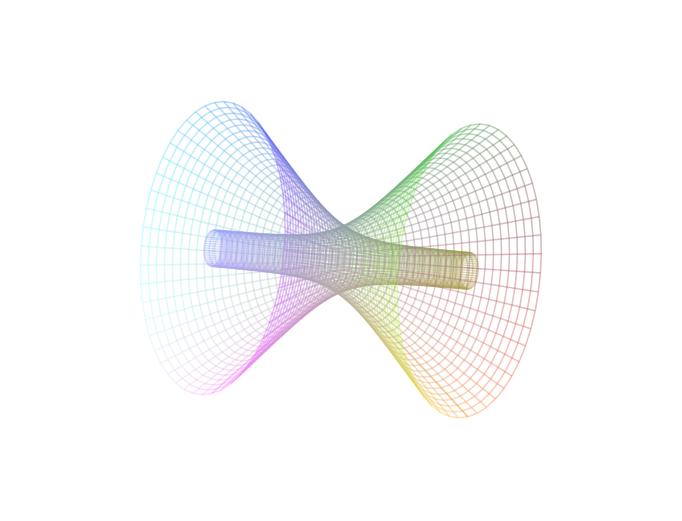}
	\caption{Interpenetrating spaces in the spherical coordinates look like two intersecting funnels.}
	\label{fig:3dpicture}
\end{figure}

Both boundary conditions \eqref{sol0} and \eqref{Bound2} represent similar geometries with index permutation.    
Nevertheless the physical volume $v(Z_2)$ of the 2-dim subspace described be coordinates $Z_3 , Z_4$ is different at $Z_2 \rightarrow \pm\infty$. Therefore a nontrivial solution connecting asymptotic regions $Z_2\rightarrow +\infty$ and $Z_2\rightarrow -\infty$ should exist in analogy with well known kink solutions \cite{Raja}. 

One of the possible form of such a space is shown in Fig.\ref{fig:3dpicture}. This form is confirmed by a numerical simulation discussed below.
If an observer moves along the $Z_2-$ coordinate and intersects a point $Z_2=0$ she/he finds out an increasing of one large subspace and decreasing of another. Far from the transition at  $Z_2=0$ both of these subspaces are described by the Minkowski geometry. 

\section{Funnel-like solution}
\subsection{Direct product of the Minkowski space and compact extra space }
Let us specify a geometry of the space
and consider the space $\mathbb{V}_D$ with $D=6$ and metric in the form
\begin{equation}\label{int2}
ds^2 =e^{2\alpha (u)}dt^2 - du^2 - e^{2\beta_1(u)}G_{1,ab}dy^ady^b - e^{2\beta_2(u)}G_{2,mn}dz^mdz^n
\end{equation}
where $-\infty <u< \infty$. There are three independent functions - $\beta_1 (u)$, $\beta_2 (u)$ and the redshift function $\alpha (u)$. The variable $u$ is a proper distance coordinate. The 2-dim subspaces $\mathbb{W}_{1,2}$ are described by coordinates $y_a,z_m\, (a=3,4; m=5,6)$ and represent two spheres of radius $r_1 (u) =e^{\beta_1(u)}$ and $r_2 (u) =e^{\beta_2(u)}$. 

The action is supposed contains the higher order derivatives of metric in the form
\begin{eqnarray}\label{act1}
&& S=\frac{m_D ^{4}}{2}\int d^6 Z \sqrt{|{G}|}\left[F(R) + c_1 R_{AB}R^{AB} \right]. \\
&& F(R)= R+cR^2-2\Lambda
\end{eqnarray}
Here $c,c_1$ and $\Lambda$ are physical parameters of order $m_D$.
By common views, higher order curvature terms appear due to quantum corrections, and it seems natural to include the Ricci tensor squared $ R_{AB}R^{AB}$ and the term $\sim R^2$ on equal footing. 

To make speculation as clear as possible suppose that
the space  $\mathbb{V}_D$ represents direct product of the 4-dim "large"\, space with coordinates $t,u,y^1,y^2$ and 2-dim sphere of radius $e^{\beta_2(u)}$, see \eqref{int2}. Due to the extremal smallness of the cosmological constant, we neglect its value, $\Lambda_0 =0$. As is shown below this approximation imposes a condition to the Lagrangian parameters $c,c_1$ and $\Lambda$.


It is well known that the F(R) theory can be cast in the form of Einstein-Hilbert theory with a potential for the effective scalar-field degree of freedom \cite{Barrow,Maeda,Magnano} which strongly facilitate an analysis.
Another method for the same purpose is developed in \cite{BRu}. Both of these methods include the conformal transformation which holds only if $F'(R)\neq 0$. This condition has a profound basis. Indeed, as was shown in \cite{Nariai,Gurovich}, theories of F(R) gravity are unstable at the hypersurface $F'(R)=0$.

To proceed, let us use the method of slow motion \cite{BRu}. More definitely, consider the limit
\begin{equation}
R_{(4)} \ll R_{(2)}
\end{equation}
and assume that the metric tensor $g_{AB}$ varies slowly with the coordinate $u$.
After some calculations \cite{BOOK, BRu} we obtain the effective action in the Einstein frame
\begin{eqnarray}
&&S_{eff}=\frac{v_2}{2}\int d^4 x(sign F')\left[R_{(4)} +\frac{k(\phi)}{2}\partial_{\mu}\phi \partial^{\mu}\phi -V(\phi)\right] \\
&&k(\phi)=\frac{1}{\phi}\left[3\phi^2\left(\frac{F''}{F}\right)^2 - 2\phi \left(\frac{F''}{F}\right) +2 \right] \\
&& V(\phi)=-sign (1+2c\phi)\frac{1}{2}\frac{ |\phi |[(c+c_1/2)\phi^2 +\phi -2\Lambda]}{(1+2c\phi)^2}.
\end{eqnarray}
Here and in the following $m_D = 1$ and the Planck mass $M_{Pl}=\sqrt{v_2}$.  $v_2$ is the volume of 2-dim sphere of unit radius. 

\begin{figure}
	\centering
	\includegraphics[width=0.6\linewidth]{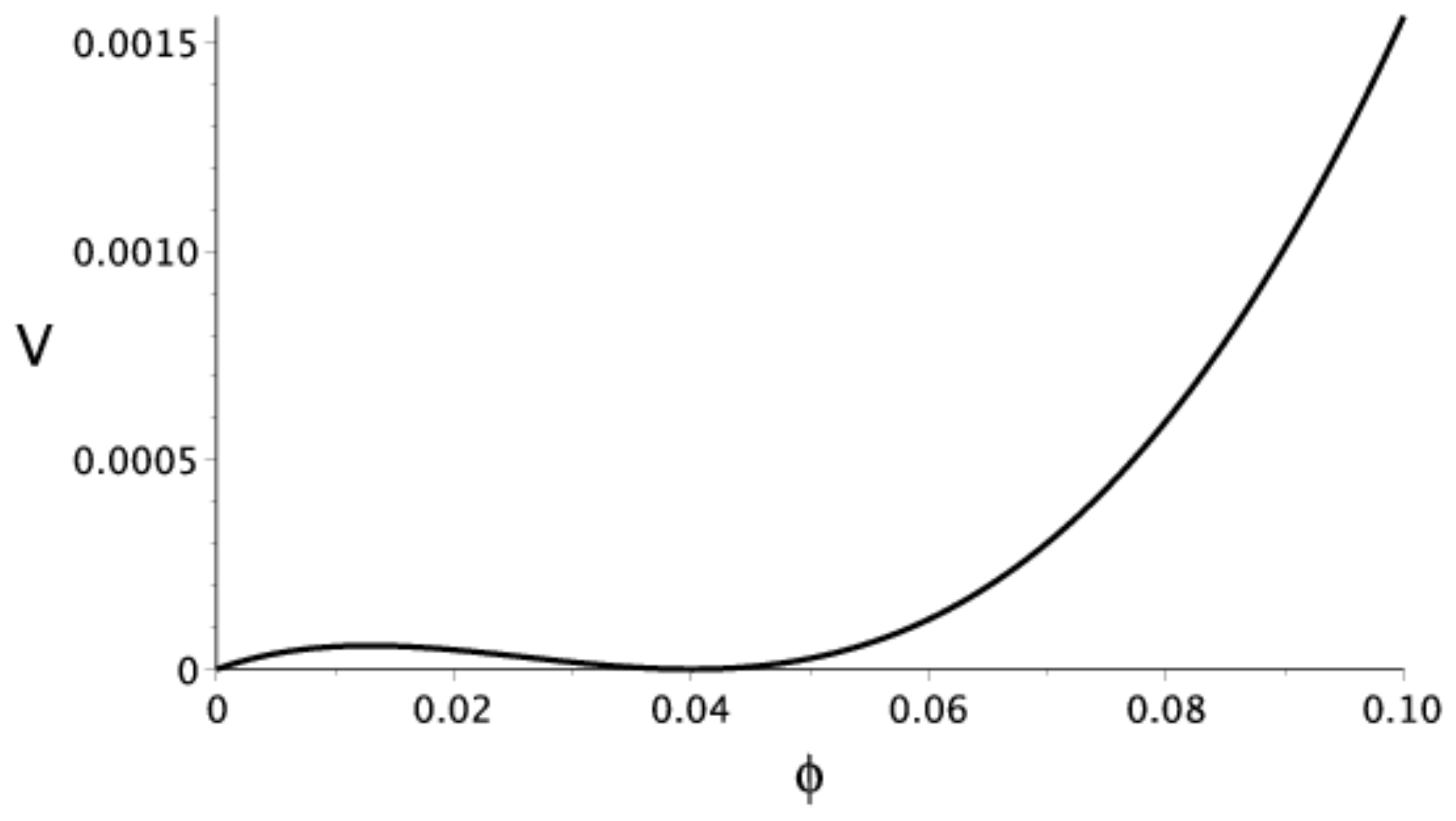}
	\caption{ The plot of the potential density $V$. The Lagrangian parameters $c=5,\Lambda=0.01,c_1 =-27$. }
	\label{V}
\end{figure}

The potential density $V(\phi)$ represented in Fig.\ref{V} depends on the scalar field which is connected to the Ricci scalar $R_{(2)}$ of the extra space, $\phi(u)\equiv R_{(2)}=2e^{-2\beta(u)}$. The presence of the potential minimum indicates stationarity of extra space of constant curvature.

Necessary conditions for the cosmological constant be equal zero have the form
\begin{equation}
 V(\phi_M)=0,\quad  V'(\phi_M)=0.
\end{equation}
These equations fix the field
\begin{equation}
\phi_M = 4\Lambda
\end{equation}
and give the connection between the physical parameters
\begin{equation}
\Lambda = \frac{-1}{4(2c+c_1)}.
\end{equation}
The radius $r_0$ of the extra space is expressed in the form
\begin{equation}
r_0 = e^{\beta_M}=\sqrt{\frac{2}{\phi_M}} = \frac{1}{\sqrt{2\Lambda}}
\end{equation}

As the result we have the 6-dim space as the direct product of the 4-dim Minkowski space and static 2-dim extra space with metric functions
\begin{equation}\label{sol1}
\alpha(u ) =0,\, \beta_1 (u) = ln|u|, \beta_2 (u)= ln\left(r_0\right)
\end{equation} 
in the interval \eqref{int2}. One can easily check that 
\begin{equation}
F'(R)=1+2c\phi_M =1+8c\Lambda
\end{equation}
is not equal zero for the numerical values of parameters listed in the capture of Fig. \ref{V}.
 Now let us come back to more complex metrics with different boundary conditions. 

\subsection{Funnel solution}

The choice of boundary conditions is the key point for a search of new stationary metrics. Let the first of them (at $u\rightarrow +\infty$.) coincides with stationary solution \eqref{sol1}
The second boundary condition (at $u\rightarrow -\infty$) in the form
\begin{equation}\label{sol2}
\alpha =0,\, \beta_2 = ln|u|, \beta_1 = ln\left(r_0\right), \quad u\rightarrow -\infty
\end{equation}
is obtained by the index substitution $1\longleftrightarrow 2$. Metric \eqref{sol2} relates to the Minkowski space with coordinates $t,u,z^1,z^2$.   A kink-like solution should fit two essentially different stationary  metrics at  $u\rightarrow \pm \infty$.  

In this paper, numerical solution was found by the Ritz method. To this end one should  perform minimization procedure of action \eqref{act1} on a class of trial functions. Let us choose them in the form
\begin{eqnarray}\label{trial}
&&\alpha(u,w)=\frac{1}{2}\ln\left(\frac{1}{\cosh(wu)}+1\right), \\
&&r_1(u;w)=\frac{1}{2}\int_{-\infty}^{u}(\tanh(wx)+1) dx+r_{0} \nonumber\\
&&r_2(u;w)=r_1(-u;w) \nonumber
\end{eqnarray}
with free parameter $w$.  Here the definitions $r_{1,2}(u)\equiv e^{\beta_{1,2}(u)}$ are used. Asymptotic behavior
\begin{eqnarray}
&&\alpha(u\rightarrow \pm\infty;w) =0,\\ &&r_1(u\rightarrow +\infty;w)=u , \, r_1(u\rightarrow -\infty;w) = r_{0}, \nonumber \\
&&r_2(u\rightarrow -\infty;w)=u , \, r_2(u\rightarrow +\infty;w) = r_{0} \nonumber
\end{eqnarray} 
of these functions meet boundary conditions \eqref{sol1} and \eqref{sol2}.

Numerical value of the adjusted parameter $w$ is obtained by optimizing the action \eqref{act1} as the function of $w$. The Ricci scalar and the Ricci tensor squared depend on the functions $\alpha,\beta_1,\beta_2$ as follows
\begin{eqnarray}\label{Ricci}
&R=R(\alpha,\beta_1 ,\beta_2)=&2e^{-2\beta_1}+2e^{-2\beta_2}-2\alpha''-4\beta_1''-4\beta_2'' -2\alpha'^2  \nonumber \\ 
&& -6\beta_1'^2 -6\beta_2'^2 -  8\beta_1'\beta_2'
\end{eqnarray}
and

\begin{eqnarray}
&&R_{AB}R^{AB}=\sum_{i=1}^{6} (R_i^i)^2,\\
&& R_1^1 = -\alpha''-\alpha'^2, \nonumber \\
&& R_2^2 = -\alpha''-\alpha'^2-2(\beta_1'^2 +\beta_2'^2+\beta_1''+\beta_2''), \nonumber \\
&& R_3^3 =R_4^4 = e^{-2\beta_1}-2\beta_1'^2 - 2\beta_1'\beta_2' - \beta_1'' , \nonumber \\
&& R_5^5 =R_6^6 = e^{-2\beta_2}-2\beta_2'^2 - 2\beta_2'\beta_1' - \beta_2'' \nonumber
\end{eqnarray}
for metric \eqref{int2}.

\begin{figure}
\centering
\includegraphics[width=0.6\linewidth]{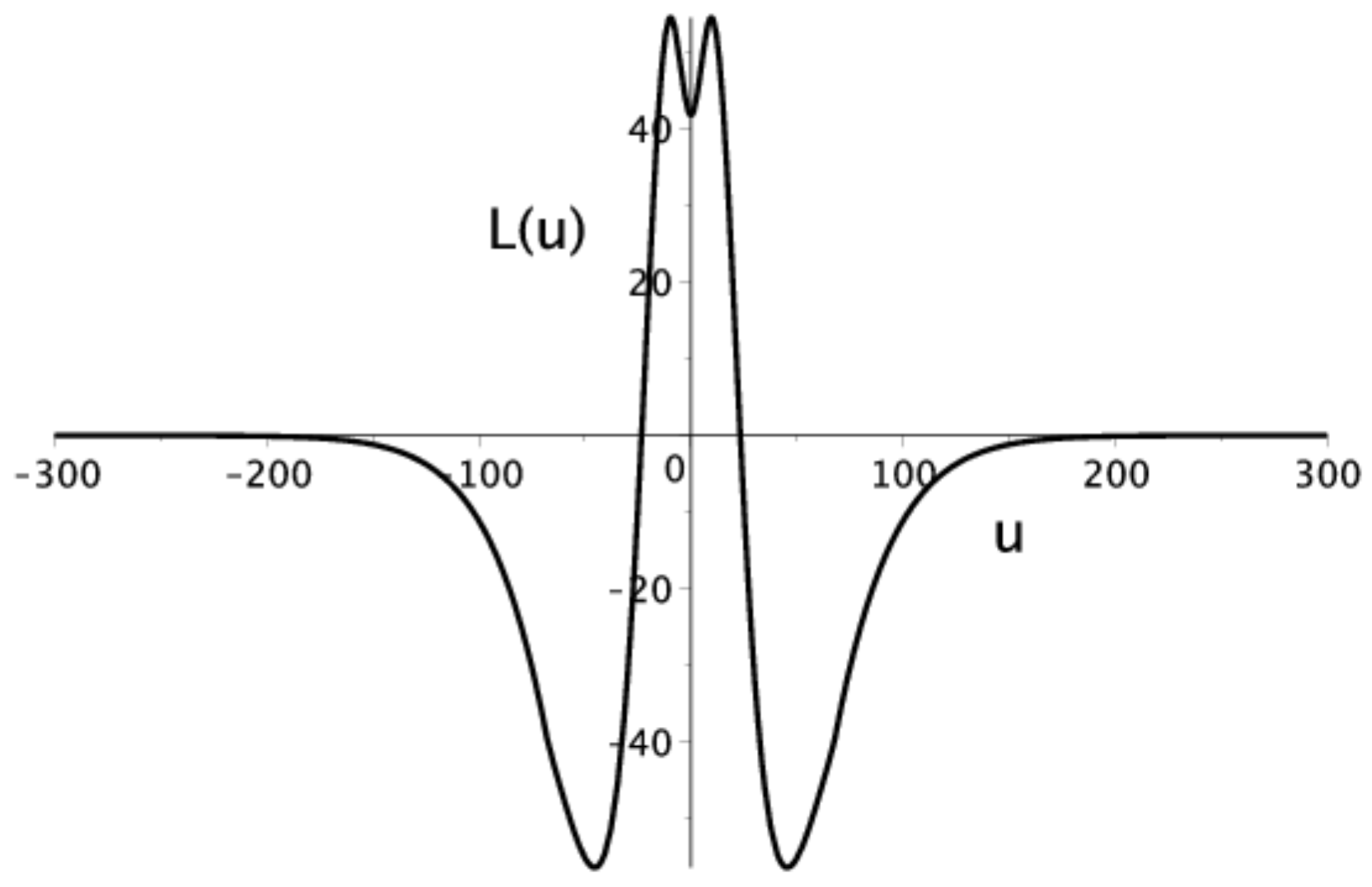}
\caption{The Lagrangian density \eqref{act1} depending on the radial coordinate $u$ for funnel solution. The Lagrangian density equals zero if the boundary conditions are equal at $u=\pm \infty$.}
\label{Lagr}
\end{figure}

\begin{figure}
	\centering
	\includegraphics[width=0.6\linewidth]{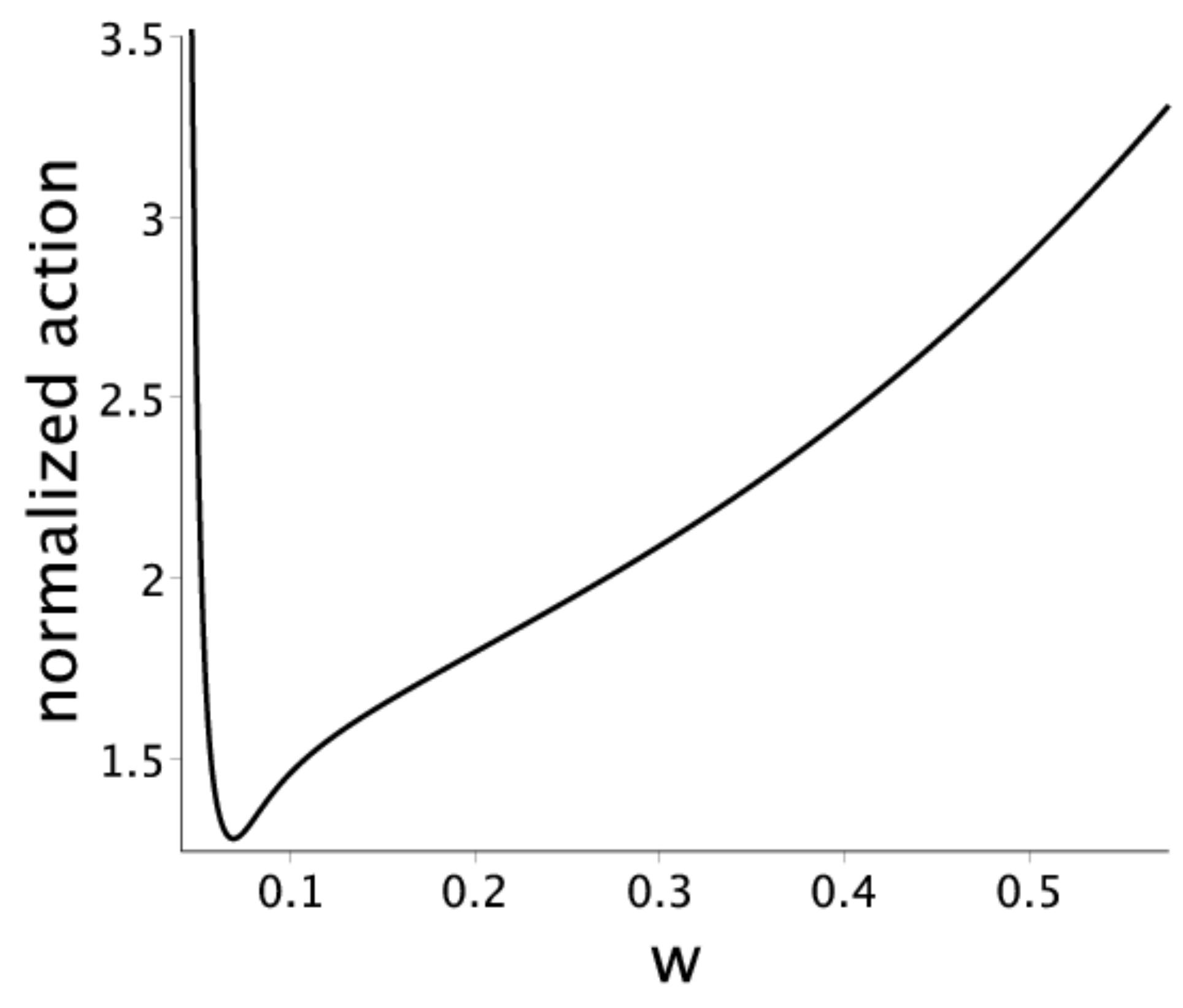}
	\caption{ The action dependence on the parameter $w$ with the minimum at $w=0.05$. The Lagrangian parameters $c=5,\Lambda=0.01,c_1 =-27$. }
	\label{action}
\end{figure}

\begin{figure}[h]
	\centering
	\includegraphics[width=0.6\linewidth]{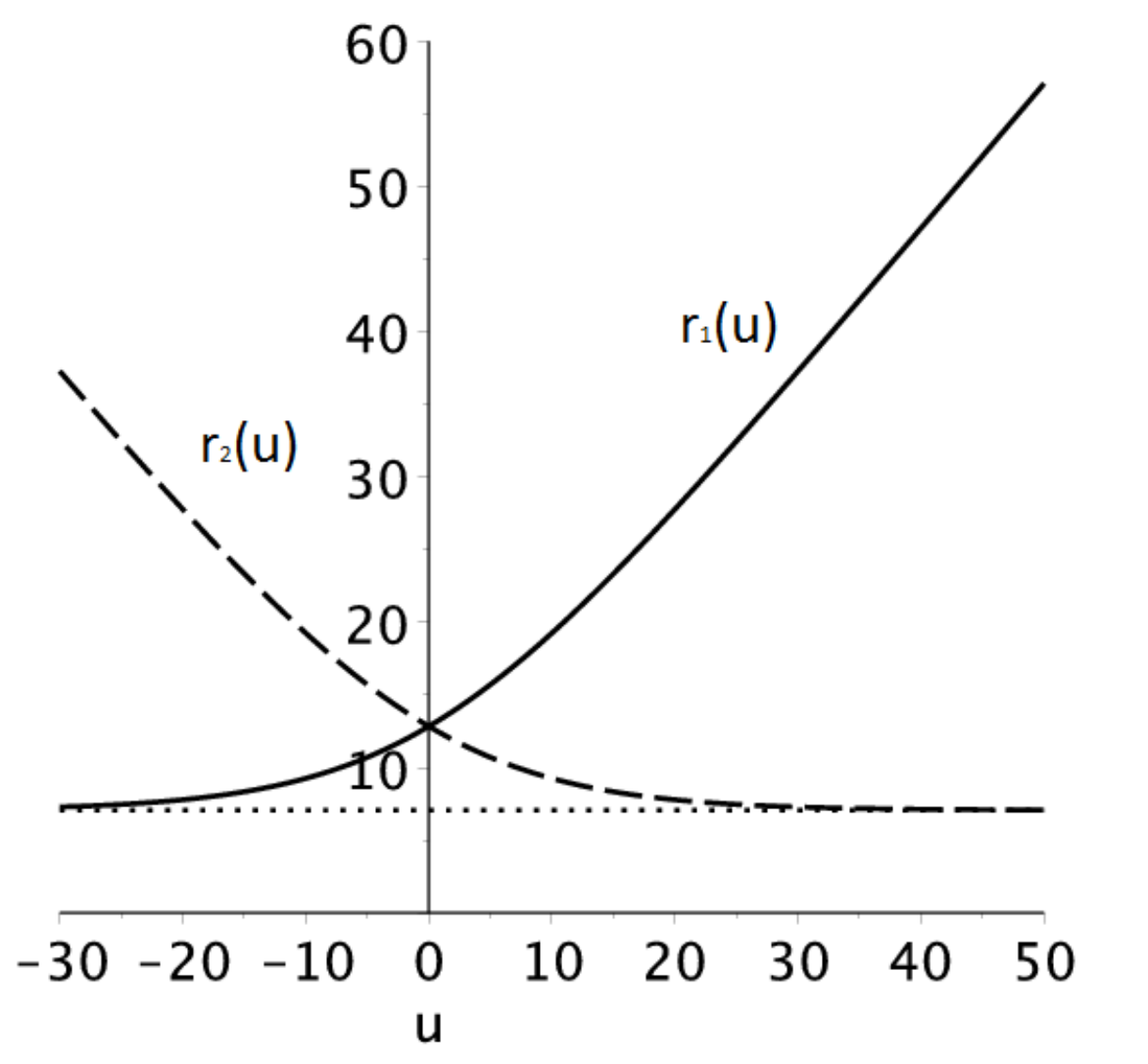}
	\caption{Two intersecting funnels as the result of numerical simulations. The size $r_{1,2}(u)$ of the extra spaces $W_{1,2}$ vs. the coordinate $u$. The horizontal line denotes the size $r_{0}$ of extra space at $u\rightarrow \pm \infty$.}
	\label{metric}
\end{figure}

The result of numerical calculation   is represented in Fig.\ref{action}. The profound minimum is attained at the point $w=w_m=0.05$ for the Lagrangian parameter listed in the figure capture. Therefore we may choose trial functions $\alpha(u,w_m), r_{1,2}(u,w_m)=\exp\{\beta_{1,2}(u,w_m)\}$ from \eqref{trial} as the approximate solution for metric \eqref{int2}. The latter is represented in Fig.\ref{metric} and Fig.\ref{fig:3dpicture}. An observer that is moving to the point $ u = 0 $ feels the increase in the size of extra space. After passing through the point $ u = 0 $, roles  of the  2-dimensional subspaces are reversed. Those which were large became small and vise versa.

Note that conditions $\beta'_1(u=0)=\beta'_2(u=0)= 0$ crucial for wormholes are not necessary now. Hence we can obtain stationary wormhole-like solutions without introducing phantom fields.

\section{Conclusion}
New solution connecting two universes is found. For an external 4-dim observer such a funnel looks similar to a spherical wormhole though its internal structure is different.  Wormhole represents two 4-dim space areas smoothly connected to each other while the solution discussed in the paper represents two 6-dim interpenetrating space areas each containing 4-dim Minkowski space and 2-dim compact extra space. 

Higher derivative gravity acting in 6-dimensional space is the basis of the study that allows to obtain the solution (funnel) without introducing a matter. Stability of the solution is maintained by the difference in asymptotic behavior at  $\pm\infty$.
 
The funnel is observed as a massive object  with a "throat"\, size smaller than $\sim 10^{-18}$cm. They could contribute to the dark matter provided that their abundance is sufficient.

\section{Acknowledgment}
The author is grateful to K. Bronnikov, V. Dokuchaev for interesting communications and to V. Berezin, K. Belotsky, A. Grobov, A. Dmitriev and A. Kirillov for helpful discussions.

This work was performed within the framework of the Center FRPP supported by MEPhI Academic Excellence Project (contract ¹ 02.à03.21.0005, 27.08.2013) and was supported by the Ministry of Education and Science of the Russian Federation, Project No.~3.472.2014/K.


\begin{thebibliography}{99}
\bibitem{Bron} K.A. Bronnikov, Acta Phys. Pol., B \textbf{4}, 251(1973).


\bibitem{ArmendarizPicon}
C.~Armendariz-Picon,
Phys.\ Rev.\ D {\bf 65} (2002) 104010
[gr-qc/0201027].

\bibitem{deAlbuquerque}
L.~C.~de Albuquerque, P.~Teotonio-Sobrinho and S.~Vaidya,
JHEP {\bf 0410} (2004) 024
[hep-th/0407041].

\bibitem{DeBenedictis}
A.~DeBenedictis and A.~Das,
Nucl.\ Phys.\ B {\bf 653} (2003) 279
[gr-qc/0207077].


\bibitem{Roman} 
T.~A.~Roman,
Phys.\ Rev.\ D {\bf 47}, 1370 (1993)
[gr-qc/9211012]. 

\bibitem{Krishnan}
C.~Krishnan, S.~Paban and M.~Zanic,
JHEP {\bf 0505} (2005) 045
[hep-th/0503025].

\bibitem{Sushkov}
S.~V.~Sushkov and S.~W.~Kim,
Gen.\ Rel.\ Grav.\  {\bf 36} (2004) 1671
[gr-qc/0404037].


\bibitem{Dzhun}
Vladimir D. Dzhunushaliev, D. Singleton, Phys.Rev. D59 (1999) 064018 [gr-qc/9807086].


\bibitem{Star80} A. A. Starobinsky, Phys. Lett. B 91, 99 (1980)



\bibitem{Amendola} L. Amendola, R. Gannouji, D. Polarski and S. Tsujikawa, Phys. Rev. D \textbf{75},  083504 (2007).

\bibitem{Starob1} A. A. Starobinsky, JETP Lett. \textbf{86}, 157 (2007).

\bibitem{Odin} G. Cognola, E. Elizalde, S. Nojiri, S. D. Odintsov, L. Sebastiani and S. Zerbini, Phys. Rev.D \textbf{77},046009  (2008).


\bibitem{Barrow}J.D. Barrow and S. Cotsakis, Phys. Lett. B214, (1988), 515 - 518.

\bibitem{Maeda}
K. Maeda, Phys. Rev. D39, (1989), 3159 - 3162.


\bibitem{Magnano}
G. Magnano and L.M. Sokolowski, Phys. Rev. D50, (1994), 5039 - 5059, gr-qc/9312008;


\bibitem{BRu}
K.~A. {Bronnikov} and S.~G. {Rubin}.
Phys. Rev. D \textbf{73}, 124019 (2006).

\bibitem{BOOK} Kirill A Bronnikov and Sergey G Rubin "BLACK HOLES, COSMOLOGY AND EXTRA DIMENSIONS", World Scientific Publishing Co. Pte. Ltd., 2012.

\bibitem{Nariai}H. Nariai, Progr. Theor. Phys. 49, 165 (1973)

\bibitem{Gurovich} V. Ts. Gurovich and A. A. Starobinsky, Sov. Phys. - JETP 50, 844 (1979).

\bibitem{DiCriscienzo}
R.~Di Criscienzo, R.~Myrzakulov and L.~Sebastiani,
Class.\ Quant.\ Grav.\  {\bf 30} (2013) 235013
[arXiv:1306.4750 [gr-qc]].

\bibitem{ab(t)1} E.A. Leon, R. Nunez-Lopez, A. Lipovka, J.A. Nieto, 
Mod.Phys.Lett. A26 (2011) 805-814 [arXiv:1012.3556].

\bibitem{KKReview} T. Appelquist, A. Chodos and P. G. O. Freund, "Modern Kaluza-Klein Theories," (Frontiers in Physics, Vol 65).

\bibitem{Raja} R. Rajaraman, Solitons and Instantons, North-Holland Publishing Company, 1982.
\end{thebibliography}
\end{document}